\documentclass[nonacm]{acmart}

\usepackage{color}

\begin{document}

\title[Internet-human infrastructures]{Internet-human infrastructures: Lessons from Havana's StreetNet}

\author{Abigail Z.\ Jacobs}
\authornote{Both authors contributed equally to this research.} \email{azjacobs@umich.edu}
\author{Michaelanne Dye}
\authornotemark[1]
\email{mmtd@umich.edu}
\affiliation{
   \institution{University of Michigan}
}

\renewcommand{\shortauthors}{A.\ Z.\ Jacobs and M.\ Dye}

\begin{abstract}

We propose a mixed-methods approach to understanding the \emph{human infrastructure} underlying StreetNet (SNET), a distributed, community-run intranet that serves as the primary 'Internet' in Havana, Cuba.
We bridge ethnographic studies and the study of social networks and organizations to understand the way that power is embedded in the structure of Havana's SNET.
By quantitatively and qualitatively unpacking the human infrastructure of SNET, this work reveals how distributed infrastructure necessarily embeds the structural aspects of inequality distributed within that infrastructure.
While traditional technical measurements of networks reflect the social, organizational, spatial, and technical constraints that shape the resulting network, ethnographies can help uncover the texture and role of these hidden supporting relationships.
By merging these perspectives, this work contributes to our understanding of network roles in growing and maintaining distributed infrastructures, revealing new approaches to understanding larger, more complex Internet-human infrastructures---including the Internet and the WWW. 
\end{abstract}


\begin{teaserfigure}
  \includegraphics[width=\textwidth]{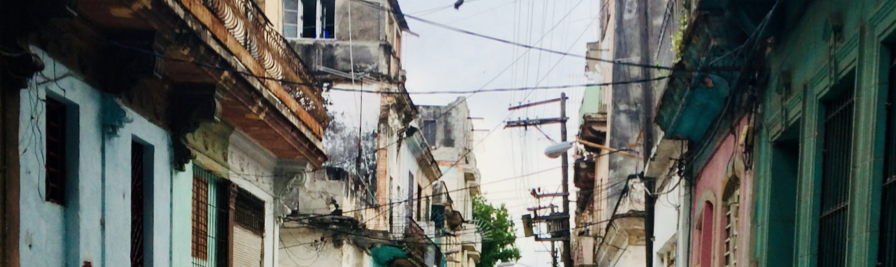}
  \caption{While few Cubans have access to the WWW, the plurality of Cubans use an isolated, community-run \emph{ad hoc} network known as StreetNet (SNET). 
  The physical structure and connections of this network are salient, in part due to ongoing community practices of maintenance and care, and its visibility. Shown here, cables connecting SNET nodes are strung across and between residences in Havana, Cuba. Photo credit: Michaelanne Dye}
  \label{fig:wires}
\end{teaserfigure}

\maketitle


\section{Distributed human infrastructures}

Cuba has one of the lowest Internet penetration rates worldwide  
but, in the capital city of Havana, thousands of volunteers sustain arguably ``the largest isolated community-driven network in existence" \cite{pujol2017initial}---StreetNet, or SNET. SNET is an informal distributed network with highly visible infrastructure: physical cables between homes can be visible from the street (Fig.\ \ref{fig:wires}). Due to the specific context of SNET, we argue that it provides a compelling case from which to explore mixed-method approaches to studying complex sociotechnical networks. 

Despite traditional perceptions of infrastructures as invisible and seamless \cite{Star1994}, scholars have highlighted the precarity and instability of infrastructures, particularly in the global South \cite{Larkin2013,Jackson2014,Sambasivan2010}. Lee et al.\ define \emph{human infrastructure} as ``the arrangements of organizations and actors that must be brought into alignment in order for work to be accomplished'' \cite{Lee2006}, and others have argued for more work considering human infrastructure, incorporating both social and technological aspects of information infrastructures \cite{Sambasivan2010}. 
In SNET, the social and the technological infrastructure are highly visible.  
Users and non-users help maintain the network on a day-to-day basis: individuals lean on family and community ties to conduct frequent and occasionally time- and resource-intensive maintenance of SNET \cite{dye2019if-short}.
Maintaining social ties is necessary for day-to-day \emph{maintenance} of the physical network, and the structure of the physical network shapes social tie formation and maintenance. \looseness=-1
Furthermore, the addition of new nodes, system repair, and system administration are mediated by social ties, access to resources, connections to existing nodes, and geography. That is, SNET's \emph{growth and day-to-day functioning} is mediated by social capital and constraints. In turn, this has spillover effects, shaping access to SNET as well as its governance. 
Understanding human infrastructure---here, in the case of SNET---requires a mixed-methods approach, drawing from ethnography, social networks, and empirical studies of the Internet and Web. 
\looseness=-1


\section{Studying distributed human infrastructures} 

Here, we briefly review several academic traditions that can address different aspects of human infrastructures.\looseness=-1

\subsubsection*{Internet measurement}

Designing a more decentralized, resilient, and invisible infrastructure has been a goal in the design of the Internet (e.g., \cite{clark2005tussle,clark2018designing,feamster2017and}).
This emphasis presumes a robust set of technologies and opportunities for access, 
while masking how structural, organizational, and political constraints shape the development and maintenance of the Internet \cite{edmundson2016first,edmundson2018nation,holland2019measuring,leyba2019borders}. 
In the case of Cuba and other states that are poorly connected to the Internet (such as Reunion Island~\cite{noordally2016analysis}), there is an opportunity to study the `edges' of the Internet \cite{bischof2015and}. Measurement along this boundary of the Internet affords an opportunity to understand how geopolitical processes have shaped the structure of the Internet \cite{chen2009sidewalk,edmundson2018nation,leyba2019borders}.  
Studying the Internet as a technology in isolation of its broader human (and organizational) infrastructure can overlook important constraints for growth, maintenance, and access. Abbate, for instance, argues that ``one limitation of defining the Internet as a large technological system or infrastructure is that this tends to frame the Internet as a channel for transmitting data, rather than as a field of social practice"---also noting that ``a systems approach also privileges the role of system builders over users"
\cite{abbate2017and}. 
In such a large system, these processes of human infrastructure can be hidden, only becoming visible at the seams (in case of breakdown) or at the edges (in the case of poorly connected regions). 
As a first step to studying larger, more complex systems, SNET provides an opportunity to explore the role of human infrastructures.
The scale and visibility of SNET allow us to study how human infrastructure has, and could, shape critical infrastructures.\looseness=-1

\subsubsection*{Economic sociology \& social networks}

Economic sociology highlights 
how markets are embedded in social ties.
This notion of \emph{embeddedness} reveals how the structure and exchange of resources (including information) are mediated by underlying social structures \cite{granovetter1992problems,granovetter1985economic,krippner2004polanyi}.  
The theory and methods of social networks give researchers a tool to 
map this embeddedness, describing the social structure of individuals and how this connects to individual outcomes and different environments \cite{granovetter1985economic,granovetter1992problems,granovetter2005impact,krippner2004polanyi}.
While economic sociology gives us theory and network perspectives, the field has largely overlooked how the
``hidden architectures" \cite{gonzalez2017decoding} of infrastructure shape and are shaped by social structure \cite{pinch2008living}. 
The visibility of the human infrastructure of SNET allows us to understand embeddedness in modern infrastructures.
\looseness=-1

\subsubsection*{Ethnography and the human infrastructure of the internet(s)}

 Ethnography acts as both a rigorous method and a way of knowing \cite{Marcus1998}, serving as a critical means to study and document social, cultural, and technical processes. Ethnography  branched out of anthropology into the fields of Science and Technology Studies (STS), Computer Supported Cooperative Work (CSCW), and Human Computer Interaction (HCI), among others. In studying the interplay between people and technology, ethnography is invaluable in providing an in-depth perspective of the lived conceptualizations of technoculture. In a network as complex and \textit{seamful} as SNET, ethnography provides a valuable lens for studying crucial connections in the network. Prior work in STS and CSCW has drawn on ethnography to study social and information infrastructures 
 in order to foreground the ``the social system of human actors, activities, spaces, networks, and goals'' that often remain in the background \cite{Lee2006,Sambasivan2010}. 
 Dye expands upon this work to define \emph{collaborative configuration} as the ongoing actions, large and small, that people undertake to sustain the social, technical, and physical aspects of sociotechnical systems \cite{dye2019vamos}.\looseness=-1

Dye uncovered elements of the human infrastructure and collaborative configurations in SNET through ethnographic methods conducted during fieldwork trips throughout 2015-2017. During this time, she conducted in-depth, semi-structured interviews with 46 individuals, which included administrators of varying levels, users, and users' family members and neighbors who were involved in helping maintain the equipment. She also conducted participant observation among seven different nodes located in neighborhoods across Havana. Dye spent hundreds of hours observing and participating in SNET-related activities, including participating on the network and undertaking the work to maintain both physical and social connections crucial to the maintenance of the network. Through this work, she was able to uncover elements of the SNET infrastructure that had not been visible in previous research \cite{pujol2017initial}. 

\section{Mixed methods approaches for studying human infrastructure}

Each of these perspectives attends to different levels and aspects of distributed human infrastructures.
From ethnography, we can observe thick contextual relationships in the social practices around infrastructure, and how these practices are shaped by their environment. 
Economic sociology and social networks provide theory and tools to understand context---how the technical and administrative aspects of the network are embedded in social relationships \cite{granovetter1992problems}. However, infrastructure is often overlooked in economic sociology \cite{pinch2008living}. 
Empirical studies of the Internet provide insight into patterns of behavior, patterns of connectivity, and the role of geographic context on Internet connectivity. However, emphasis on the invisibility, resilience, and egalitarian nature of distributed systems may overlook important challenges to governance, growth, and maintenance. \looseness=-1
We then must take a \emph{mixed methods} approach to 
uncover the social and organizational processes that shape the development and growth of critical infrastructures. 
In this section, we apply a mixed-methods approach to the case of SNET, drawing on ongoing ethnographic work conducted by Dye. We highlight several empirical opportunities and challenges in this space that are only made visible at the intersection of these methods. \looseness=-1 

\subsubsection*{Seamfulness as empirical opportunity}

Borrowing the language of seams, or the gaps between systems \cite{weiser1991computer}, Vertesi highlights \emph{seamfulness}, or how infrastructures lay in messy overlaps with other systems: instead of presenting stable nodes, ``seams suggest that there are many possible ways to patch multiple systems together into local alignment'' (\cite{Vertesi2014}:6).
The visibility of these seams in SNET provides an opportunity to study the role of human infrastructures. As the physical and social network may undergo changes independent of one another, we can observe how each system adapts to new configurations: even without observing the full physical or  social network,  these perturbations still provide meaningful insight into how human infrastructure shapes and is shaped by the network. 


\subsubsection*{Promises of decentralization}

Decentralized, distributed infrastructures, such as SNET or the Internet, are 
often designed to reflect values of robust, egalitarian, seamless infrastructure \cite{brown2010should,clark2018designing,pujol2017initial}. %
However, inherently networked systems are necessarily limited by physical, geographic, social, and organizational constraints. 
The resulting networks will then necessarily encode and entrench patterns of inequality.
The diverse roles of users and non-users in SNET; the structure of SNET; the limited access to capital, information, and resources; and political and cultural constraints on behavior all point to a need to bridge ethnographic studies, internet measurement, and social networks to understand its embeddedness, past and future.
Mapping the social structure of human infrastructure can reveal how inequality 
shapes (and is shaped by) the design, growth, and maintenance of decentralized, distributed systems.
\looseness=-1

\subsubsection*{Brokerage in the physical and social network}
The physical layout of SNET naturally creates \emph{brokers}, where individuals can play an outsize role in the transfer of information and resources \cite{stovel2012brokerage}. 
These brokers can be switch operators located at important physical positions in the network, connecting two separate regions, or serve in prominent administrative roles, enforcing disconnections and suspensions for users.
The growth and maintenance of SNET is then necessarily shaped by those brokers due to their network position \cite{granovetter1985economic}.
In particular, we find that brokers can \emph{exploit} their position in the physical network by extracting rent,
and brokers in more removed, less densely-populated locations (city outskirts) are better able and more likely to exploit their position. Their positions are well maintained by the high resource-, time-, and social cost of changing  the network.
On the other hand, we also find that brokers in more removed, less densely-populated locations also are made more \emph{accountable} for their maintenance and governance practices: as their identity is more likely to be known within the community, individuals are less able to exploit their position.
These conflicting mechanisms are driven by different social and physical features of the network, highlighting different incentives for forming and maintaining social ties in the network \cite{stovel2011stabilizing}.
Understanding the social processes at play can only be done by attending to \emph{both} the physical network and the human infrastructure, and ethnographies are needed to understand the network contexts. 
\looseness=-1

\subsubsection*{Ethical challenges}

Particularly in contexts such as Cuba's, it is crucial that researchers recognize and address potential risks to participants from conducting and publishing this type of work. SNET is not officially sanctioned by the government and could be considered illegal, therefore extra care must be taken to ensure participants' anonymity. Further, by bringing more attention to the network through publication, we may be bringing increased scrutiny from the Cuban government. However, participants from Dye's work assured her that the Cuban government is aware of SNET and has not attempted to disrupt it. SNET occupies a legal ``gray'' area along with other unofficial networks in Cuba (like El Paquete Semanal \cite{dye2018paquete}). In addition to recent academic work \cite{pujol2017initial}, SNET has received international media attention \cite{Estes2015,Martinez2017,Crecente2017}, with members of the network appearing on camera or in photographs \cite{Estes2015}. In addition to anonymizing participants' names, we also recommend using pseudonyms for network nodes and neighborhoods to further protect participants' privacy.\looseness=-1 

\section{Looking forward} 

The social and geographic aspects of infrastructure are particularly salient in the context of SNET, but exist across other contexts and scales of Internet-human infrastructures. Ethnographies can help uncover the texture and role of these hidden supporting relationships, while traditional technical measurements of networks reflect the social, organizational, spatial, and technical constraints that shape the resulting network. 
 However, there currently is both an \emph{empirical and theoretical} gap between these observed, deep individual contexts and system-level structures. 
We seek to bridge this gap by bringing a mixed-methods perspective to understanding Cuba's StreetNet, a critically important infrastructure. This can pave the way to study the governance, formation, and maintenance of the Web and Internet across a range of social, cultural, and political contexts.\looseness=-1

\bibliographystyle{ACM-Reference-Format}
\bibliography{snetrefs}

\end{document}